\documentclass[%
reprint,
superscriptaddress,
nofootinbib,
amsmath,amssymb,
aps,
pre,
]{revtex4-2}

\usepackage{gensymb,units,textcomp,color,leftidx,xspace,epsfig,units,bm,graphicx,lipsum,tabularx}
\usepackage[acronym]{glossaries}
\usepackage[caption=false]{subfig}

\newcolumntype{Y}{>{\hsize=.6\hsize}X}
\newcolumntype{Z}{>{\hsize=.5\hsize}X}
\newacronym{CCD}{CCD}{Charge Coupled Device}
\newcommand{\CCD}{\gls{CCD}\xspace}
\newacronym{IPD}{IPD}{ionization potential depression}
\newcommand{\IPD}{\gls{IPD}\xspace}
\newacronym{LCLS}{LCLS}{Linac Coherent Light Source}
\newcommand{\LCLS}{\gls{LCLS}\xspace}
\newacronym{FEL}{FEL}{free-electron-laser}
\newcommand{\FEL}{\gls{FEL}\xspace}
\newcommand{\LTE}{LTE\xspace}
\newacronym{LANL}{LANL}{Los Alamos National Laboratory}
\newcommand{\LANL}{\gls{LANL}\xspace}
\newacronym{SXR}{SXR}{Soft X-Ray}
\newcommand{\SXR}{\gls{SXR}\xspace}
\newacronym{EK}{EK}{Ecker-Kr\"oll}
\newcommand{\EK}{\gls{EK}\xspace}
\newacronym{SP}{SP}{Stewart-Pyatt}
\newcommand{\SP}{\gls{SP}\xspace}
\newacronym{IS}{IS}{Ion-Sphere}

\newacronym{DFT}{DFT}{Density Functional Theory}
\newcommand{\DFT}{\gls{DFT}\xspace}
\newacronym{SASE}{SASE}{self-amplified spontaneous emission}
\newcommand{\SASE}{\gls{SASE}\xspace}

\newcommand{\hea}{He$_{\alpha}$\xspace}

\newcommand{\Oxford}{Department of Physics, Clarendon Laboratory, University of Oxford, Parks Road, Oxford OX1 3PU, UK}
\newcommand{\UVa}{Departamento de Física Teórica Atómica y Óptica, Universidad de Valladolid, 47011 Valladolid, Spain}
\newcommand{\EuXFEL}{European XFEL GmbH, Holzkoppel 4, Schenefeld, 22869, Germany}
\newcommand{\CLF}{Central Laser Facility, STFC Rutherford Appleton Laboratory, Didcot OX11 0QX, UK}
\newcommand{\ICL}{Plasma Physics Group, The Blackett Laboratory, Imperial College London, Prince Consort Road, London, SW7 2AZ, UK}
\newcommand{\SLAC}{Linac Coherent Light Source, SLAC National Accelerator Laboratory, Menlo Park, CA 94025, USA}
\newcommand{\IoP}{Department of Radiation and Chemical Physics, Institute of Physics, Czech Academy of Sciences, Na Slovance 2, 182 21 Prague 8, Czech Republic}

\begin{document}

\title{Dielectronic satellite emission from a solid-density Mg plasma: relationship to models of ionisation potential depression}
\author{G. Pérez-Callejo}
    \email{gabriel.perez.callejo@uva.es}
    \affiliation{\UVa}
\author{T. Gawne}
    \affiliation{\Oxford}
\author{T. R. Preston}
    \affiliation{\EuXFEL}
\author{P. Hollebon}
    \affiliation{\Oxford}
\author{O. S. Humphries}
    \affiliation{\EuXFEL}
\author{H.-K. Chung}
    \affiliation{Korea Institute of Fusion Energy (KFE), Daejeon, 34133, South Korea}
\author{G. L. Dakovski}
    \affiliation{\SLAC}
\author{J. Krzywinski}
    \affiliation{\SLAC}
\author{M. P. Minitti}
    \affiliation{\SLAC}
\author{T. Burian}
    \affiliation{\IoP}
\author{J. Chalupský}
    \affiliation{\IoP}
\author{V. Hájková}
    \affiliation{\IoP}
\author{L. Juha}
    \affiliation{\IoP}
\author{V. Vozda}
    \affiliation{\IoP}
\author{U. Zastrau}
    \affiliation{\EuXFEL}
\author{S. M. Vinko}
    \affiliation{\Oxford}
    \affiliation{\CLF}
\author{S. J. Rose}
    \affiliation{\ICL}
    \affiliation{\Oxford}
\author{J. S. Wark}
    \email{justin.wark@physics.ox.ac.uk}
    \affiliation{\Oxford}
    
\date{\today}

\begin{abstract}
    We report on experiments where solid-density Mg plasmas are created by heating with the focused output of the Linac Coherent Light Source x-ray free-electron-laser.  We study the K-shell emission from the Helium and Lithium-like ions using Bragg crystal spectroscopy.  Observation of the dielectronic satellites in Lithium-like ions confirms that the M-shell electrons appear bound for these high charge states.  An analysis of the intensity of these satellites indicates that when modelled with an atomic-kinetics code, the ionisation potential depression model employed needs to produce depressions for these ions which lie between those predicted by the well known Stewart-Pyatt and Ecker-Kroll models.  These results are largely consistent with recent Density Functional Theory calculations.
\end{abstract}

\maketitle

\section{Introduction}
The focused output of hard x-ray \glspl{FEL}, such as the \LCLS, with peak spectral brightnesses many orders of magnitude greater than those of any synchrotron, provides a means to create hot (many hundreds of eV) plasmas at exactly solid density. Each pulse created by the \FEL can, when it is operating in \SASE mode, contain of order a few mJ of energy, which can be focused to micron-scale spots with Be lenses or Kirkpatrick-Baez mirrors~\cite{kirkpatrick1948formation}.  The short duration of the pulses (typically sub-100~fs) ensures that the x-ray energy is deposited in the solid target in the focal plane on a timescale short compared with its disassembly time.  The combination of energy, spot size, and pulse duration noted above corresponds to intensities on target of order at least 10$^{17}$ Wcm$^{-2}$. 

If the photon energy of the incoming \FEL radiation exceeds that of the K-edge of the atoms in the cold target, and the subsequent ions created, the photoionization by the \FEL results in copious K-shell core holes being created, which are subsequently filled by radiative decay from the upper levels or via the Auger process. Both the photoionized and Auger electrons are ejected into the continuum, rapidly thermalizing by losing part of their energy via collisions with the ions in the lattice (collisional ionization and three-body recombination) and with the thermal electron pool (electron-electron scattering) \cite{ciricosta2016}, heating the electrons in the system to many tens or even hundreds of eV.  The filling of the core holes created by the photoionisation typically occurs on a femtosecond timescale, short compared with the \FEL pulse duration (which is typically several tens of femtoseconds), and certainly much shorter than the target disassembly time, effectively ensuring that the resultant K-shell emission comes from the target when it is still at solid density. Thus K-shell spectroscopy of the solid density plasma produced provides detailed information on the charge states produced.

There have now been several studies of the x-ray spectra produced from solid targets in the manner described above~\cite{Vinko2012,ciricosta2012,cho2012,preston2013,Vinko2014,Vinko2015,Rackstraw2015,preston2017,ciricosta2016,ciricosta2016a,vandenBerg2018,perezcallejo2020}, which have allowed a wealth of information to be gleaned concerning various properties and processes occuring in these dense plasmas, such as opacity~\cite{preston2017}, collisional rates~\cite{Vinko2015,vandenBerg2018}, and the phenomenon of saturable absorption in the x-ray regime~\cite{Rackstraw2015}.  Of particular relevance here are those studies which have looked at \IPD~\cite{ciricosta2012,Vinko2014,ciricosta2016a}.  The lowering of the energy of the start of the continuum, or \IPD,
is a fundamental process that occurs in dense plasmas as a result of the electrostatic interactions between the atom or ion and the surrounding charged particles~\cite{ZIMMERMAN1980517,griem_1997,stewart1966lowering,ecker1963lowering}, leading to a reduction in the atomic binding energies.  A knowledge of where the continuum lies for ions has a direct impact on many important plasma processes, such as the ionization and ion charge state distribution, the equation of state, and the opacity and transport properties. 

In the work cited above, the \IPD was ascertained experimentally for a particular charge state by noting the photon energy of the \FEL for which copious K$_{\alpha}$ radiation for the relevant ion was produced, this being interpreted as the photon energy needed to create a K-shell core hole by photoionization into the continuum. This method was applied to elements of relatively low $Z$ (12-14), and relied on the fact that for the electron temperatures produced -- typically less than 200 eV -- the atomic energies and plasma temperatures are such that the vast majority of the resultant $K$-shell emission cannot be produced via thermal processes, but only due to photoionization, and thus recorded x-ray spectra, although time-integrated over each individual pulse, are effectively gated by the pulse duration. The results were compared with predictions from simple semi-classical models often employed in atomic-kinetics calculations, for example the \SP~\cite{stewart1966lowering} and \EK models~\cite{ecker1963lowering}, whose mathematical expressions are discussed later in the text.

In the first of these x-ray heating experiments, studying Al, it was found that the \EK model gave better agreement with the data \cite{ciricosta2012}. As a result of matching the K-shell binding energy, no M-shell electrons were found to be bound. In that particular experiment the authors note that the spectra from some of the highest charge states (up to Helium-like) were complicated by the presence of K$_{\beta}$ radiation from the cold solid, and in this original work no firm conclusion about the degree of \IPD for them was made.  However, it was subsequently noted that were the \EK model to be applied for the higher charge states as well, the M-shell would also not be bound~\cite{preston2013}.  This latter conclusion is at odds with the work of Hoarty and co-workers~\cite{Hoarty2013}, who observed the presence of radiation from the He-$\beta$ transition of Al in experiments using optical lasers for heating and compression, and for which the spectra could not be reconciled with the \EK model. At the same time, the \SP model was also found to be inadequate for modelling both experiments.

The difficulty of finding a single, semi-classical \IPD model that captures the pertinent physics that such an approach purports to describe is perhaps not surprising, given that the plasmas under study are dense quantum systems.  Indeed, the authors of ref. ~\cite{ciricosta2012} note in their original work that the \EK and \SP models are{\it ``ultimately both unlikely to capture fully the complex physics of atomic systems embedded within dense plasma environments over wide ranges of plasma conditions and charge states.''}  The veracity of the above statement has been given further credence by recent extensive computational studies based on \DFT~\cite{gawne2023}.  These calculations reveal that one of the main underlying issues in such dense systems is that a binary distinction between electrons that are free, and in the so-called continuum, and those that are bound to an atom or ion, cannot definitively be made -- the problem is inherently quantum mechanical in nature.

Nevertheless, owing to the manner in which the majority of standard atomic-kinetics-based calculations are constructed, their use requires that such a distinction between bound and free electrons is made and thus, at least for the foreseeable future, simple \IPD models are likely to continue to be adopted.  As a result, within the work encompassed by the \DFT calculations cited above, the use of the technique of the inverse participation ratio was employed to give a measure of the degree of boundness of the Kohn-Sham wavefunctions~\cite{gawne2023}, and hence deduce where the energy of the continuum ought best to be placed, were one forced to make such an artificial division between bound and free.  Within the above constraints and caveats, it was found that for solid density Al and Mg plasmas, for the highest charge states the most appropriate energies at which to consider the electrons to be free would correspond to a position lying somewhere between those predicted by the \SP and \EK models, although the \EK model would still give a better fit for the lower charge states.  In particular, it was noted that for Li and He-like ions, it would appear that the M-shell should be treated as being bound, and radiation from the M-shell states should be experimentally observable. In fact, the authors of \cite{gawne2023} report the direct observation of He- and Li-like Mg K$_\beta$ emission.

It is in the above context that we present here an analysis of the emission spectrum from the Helium and Lithium-like ions in an \FEL-generated solid-density, optically-thin, Mg plasma.  We specifically investigate the intensity of the dielectronic satellites, where the spectator electron lies in the $n=3$ principal quantum number, i.e. the M-shell.  Studies of the intensity of such satellites, be they due to L or M-shell spectator electrons, has a long history, and has previously been shown to be of use in determining plasma conditions in both astrophysical~\cite{gabriel1972, gabriel1979, dubau1980,bitter1979,seely1987}  and laboratory based plasmas~\cite{bombarda1985,apicella1983,rosmej1998,duston1983,woltz1991,jacobs1989,audebert1984,seely1981,hansen2020}.  

In analysing the intensity of these M-shell satellites, our main finding is that the Li-like states for which the $n=3$ level is occupied by one electron are in good thermal contact with the equivalent He-like state without the $n=3$ electron, and as a result the intensity of these satellites is sensitive to the \IPD.  Furthermore, for best agreement with the experimental data given the parameters of the \FEL, we need to place the \IPD for these charge states between the \EK and \SP limits, as advised by the \DFT studies, and to this degree the data is consistent with them.

\section{Experimental set-up}

\begin{figure*}
\centering
\subfloat[\label{fig:ExpSetUp}]{%
    \includegraphics[height=0.38\linewidth]{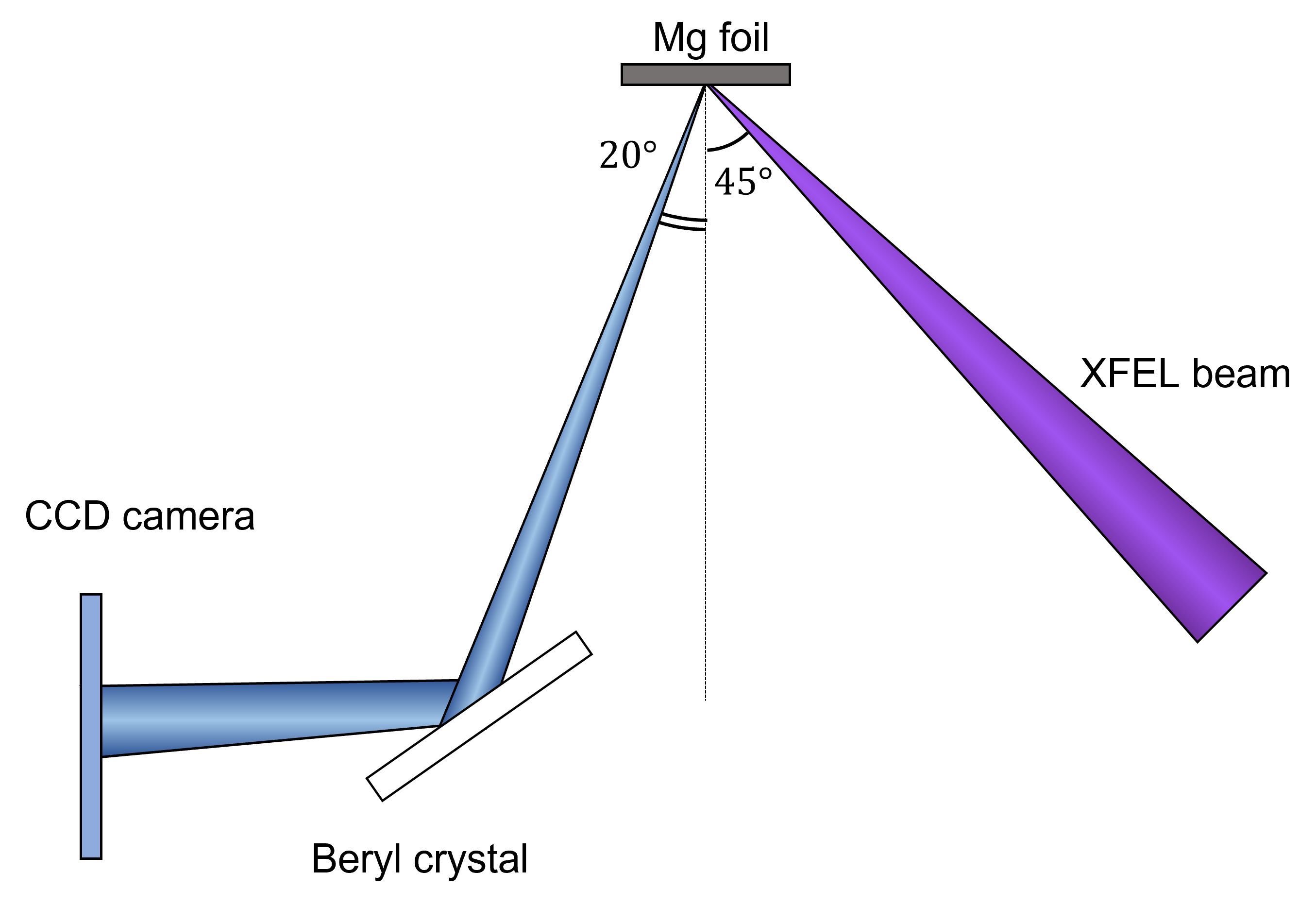}
} 
\subfloat[\label{fig:FullSpectrum}]{%
    \includegraphics[height=0.38\linewidth]{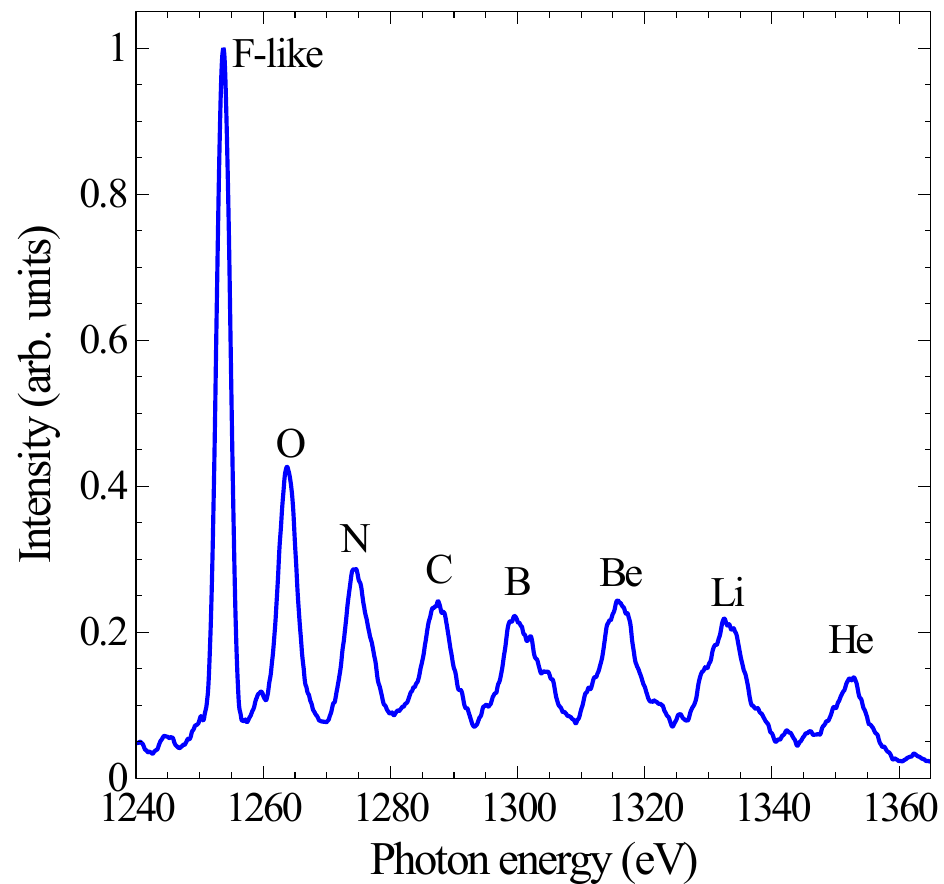}
}
\caption{(\ref{fig:ExpSetUp}) Schematic drawing of the experimental set-up. (\ref{fig:FullSpectrum}) Experimental data of Mg K-shell emission spectrum showing a whole set of emission lines from different ionization stages. The lines have been labelled according to the number of electrons of the emitting ion.}
\end{figure*}

The experiment was performed at the \SXR end station \cite{schlotter2012} of the \LCLS, using a set-up which has been thoroughly discussed in previous publications \cite{ciricosta2016, preston2017, prestonPHD, perezcallejo2020}. 

Targets consisted of \unit[54]-nm-thick Mg foils. These were irradiated using \unit[100]{fs} x-ray pulses with a photon energy of \unit[1540]{eV}. The nominal pulse energy was \unit[1.7]{mJ}, which is reduced to \unit[0.51]{mJ} after transmission through the beamline optics \cite{prestonPHD, preston2017, ciricosta2016}. The size of the focal spot on target was measured \textit{ex situ} using imprint measurements on PbI$_2$ \cite{chalupsky2010}, obtaining an effective focal spot area of $\unit[8.5]{\micro m^2}$, which corresponds to a maximum irradiance of $\sim\unit[10^{17}]{W~cm^{-2}}$.

Spectra were collected over 208 \FEL shots. Since our main interest is the shape of the \hea complex, each individual spectrum was normalized to the peak of their \hea emission. The total variance was obtained accounting for the dark noise of each pixel in the \CCD (which was in turn assigned to an energy bin), the statistical variance assuming a Poisson distribution for an individual photon observation, and the weighted variance between the set of spectra \cite{humphriesPHD}.

The targets were irradiated at a \unit[45]{\degree} angle with respect to their normal, and their x-ray emission was collected at an angle of \unit[20]{\degree} to that normal by means of a flat-crystal Bragg spectrometer. We employed a Beryl $(10\bar{1}0)$ crystal, whose lattice spacing ($2d=$\unit[15.96]{\AA}) corresponds to a diffraction angle of $\sim \unit[35]{\degree}$ for the \hea line of Mg. The diffracted X rays were then recorded with a Princeton Instruments \CCD camera. The spectral resolving power ($E/\Delta E$) of this setup was $>3000$ for all photon energies \cite{ciricosta2016a, prestonPHD}. A schematic drawing of the experimental set-up is shown in Fig. \ref{fig:ExpSetUp}.

Figure \ref{fig:FullSpectrum} shows a typical example of the K-shell emission spectra from the solid Mg plasma that was obtained in the experiment. Owing to the time-integrated nature of the measured spectra, emission from the different ionization species present in the plasma as it heats up is present in the data. The lines are labelled according to the number of bound electrons in the core of the emitting ion (the label `He' corresponds to two bound electrons, `Li' to three, and so on). As mentioned in the introduction, in this work we will focus on the emission from He-like Mg, and the associated Li-like satellites which lie in the energy range of $\unit[1340-1370]{eV}$.

\section{Modelling}
\label{sec:Model}

\subsection{Atomic kinetics}

The plasma evolution and resultant spectra were modelled using a combination of two codes: the time-dependent atomic kinetics non-\LTE (Local Thermodynamic Equilibrium) code SCFLY \cite{ciricosta2016}, and a separate, stand-alone, \LTE Saha-Boltzmann code. As we shall explain in more detail below, we perform time-dependent SCFLY simulations to determine the overall evolution of the plasma in terms of superconfigurations, and also to confirm that the system is very close to \LTE.  We then use a Saha-Boltzmann approach, with more detailed atomic physics (but now assuming \LTE) to model the spectra of the satellites for comparison with the experimental results. 

SCFLY is based upon the commonly used FLYCHK code \cite{chung2005}, adapted to treat x-ray laser problems. SCFLY is based on a superconfiguration approach -- i.e. it provides the populations for states defined solely by the number of electrons with specific principal quantum numbers.  Thus the ground state of a lithium-like ion, with two electrons in the K-shell, and one in the L-shell, is denoted (210), whereas in its first excited state, with the L-shell electron excited to the M-shell, it would be denoted (201).  It takes as its input the x-ray laser intensity as a function of time, and solves for the evolution of ground and excited-state superconfiguration populations. The electrons in the continuum are assumed to obey classical statistics, and to instantaneously thermalise to a temperature dictated by their overall energy content. This assumption has recently been validated by Ren \textit{et al.} \cite{ren2023}, who showed that for this type of plasmas, the effect of non-thermal electrons in the K-shell spectrum is negligible if the \FEL pulse is longer than $\sim\unit[30]{fs}$. In contrast with the electrons, we assume that on the timescale of typical \FEL pulses  the ions remain at room temperature throughout the calculation, given  that the timescale for electron-ion equilibriation, in terms of their temperatures, is several picoseconds~\cite{Ng1995,White2014}.

Within both the SCFLY and Saha-Boltzmann solver we take into account the degree of \IPD by considering two widely used models, namely the \EK model and the \SP model. The levels of continuum lowering predicted by these models in the high-density limit, where the \SP model is effectively equivalent to the Ion-Sphere model \cite{stewart1966lowering}, are given by \cite{preston2013}
\begin{equation}
    \Delta I_{EK} = C_{EK}\cdot \frac{(Z+1)e^2}{4\pi\varepsilon_0}\cdot \left[\frac{4\pi\left(n_e+n_i\right)}{3}\right]^{1/3}
    \label{eqn:EK}
\end{equation}
and
\begin{equation}
    \Delta I_{SP} = C_{SP}\cdot \frac{3}{2}\cdot \frac{(Z+1)e^2}{4 \pi \varepsilon_0} \cdot \left(\frac{4\pi n_e}{3\cdot (Z+1)}\right)^{1/3},
    \label{eqn:SP}
\end{equation}
where, for each model, $C_{EK}=C_{SP}=1$, $Z$ is the charge of the ion ($0$ for the neutral atom), $e$ is the electron charge, $n_e$ and $n_i$ are the electron and ion number density respectively and $\varepsilon_0$ is the electric permittivity of vacuum. Note that in order to explore models that lie between \SP and \EK we will also in what follows show results for which these two constants $C_{EK}$ and $C_{SP}$ differ from unity.

As noted in the introduction, the main result we will be studying in this work is the intensity of the lithium-like dielectronic satellites.  Within SCFLY, the transition between superconfigurations that corresponds to satellites associated with an upper state containing an L-shell electron is the (120)-(210) transition, and for M-shell satellites is the (111)-(201) transition.  At the more detailed level, taking into account the various configurations and fine-structure effects, these two  superconfigurational transitions encompass the gamut of satellites that are well known to accompany the helium-like resonance line, and which, for the L-shell satellites, are usually labelled according to the notation proposed by Gabriel~\cite{gabriel1972}.  It is these detailed transitions that are modelled by the Saha-Boltzmann code, under its \LTE assumption, and which we justify in the next section.

\begin{figure}
\includegraphics[width=0.9\linewidth]{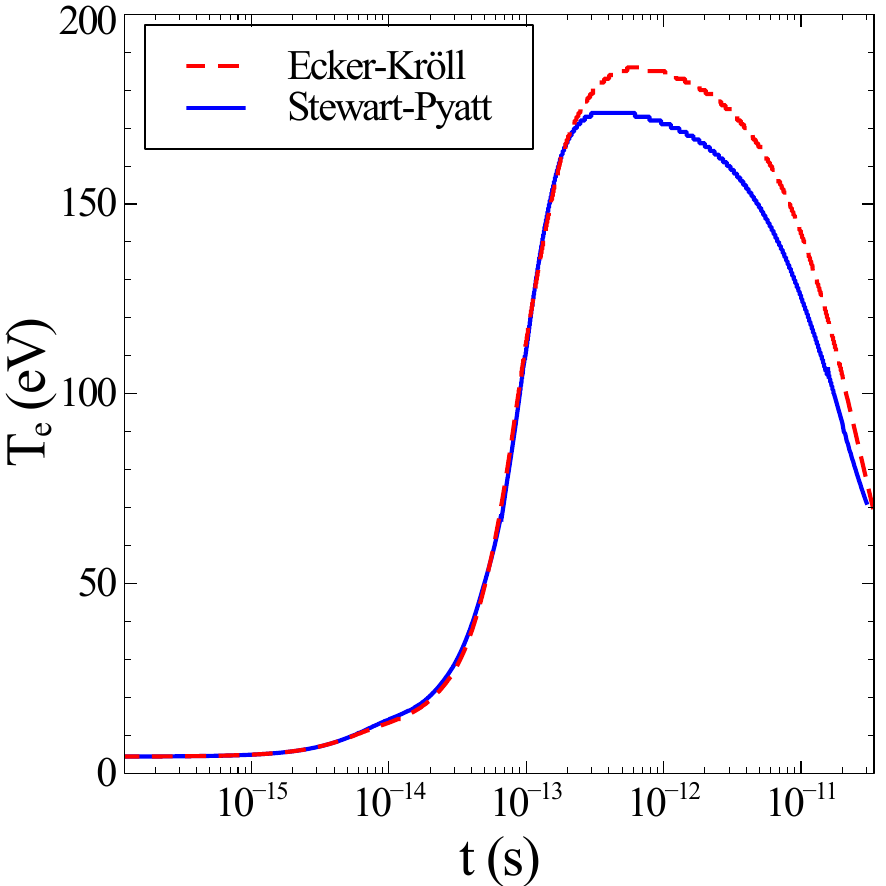}
\caption{\label{fig:Te_evolution}Temperature evolution of solid density Mg, as predicted by SCFLY for different \IPD models, upon irradiation with an \FEL pulse with irradiance $\unit[3\times10^{17}]{W~cm^{-2}}$ at \unit[1540]{eV}. The plot is shown up to \unit[10]{ps} for illustrative purposes, but it should be noted that disassembly of the target will take place on a timescale of order picoseconds.}
\end{figure}

To model the plasma kinetics, we first calculated the temperature evolution of the plasma with SCFLY. It is even at this early stage that the choice of an \IPD model affects our results, since the electron temperature of the plasma is dependent on the level of ionization and therefore on the level of continuum lowering. We show this effect in Fig. \ref{fig:Te_evolution}, where the temperature evolution for the peak laser irradiance ($\sim\unit[3\times10^{17}]{W~cm^{-2}}$) is shown for the \SP and \EK \IPD models.  In these simulations, and all that follow, we assume that the incident \FEL pulse is gaussian in time with a FWHM of \unit[100]{fs}, which peaks at a time of \unit[100]{fs}.  In the simulations shown here, which assume constant ion density, the cooling of the plasma after the \FEL pulse is only due to radiation while in practice, disassembly of the target will take place on a timescale of order picoseconds.

\begin{figure*}
    \subfloat[\label{fig:Continuum_He}]{%
    \includegraphics[width=0.49\linewidth]{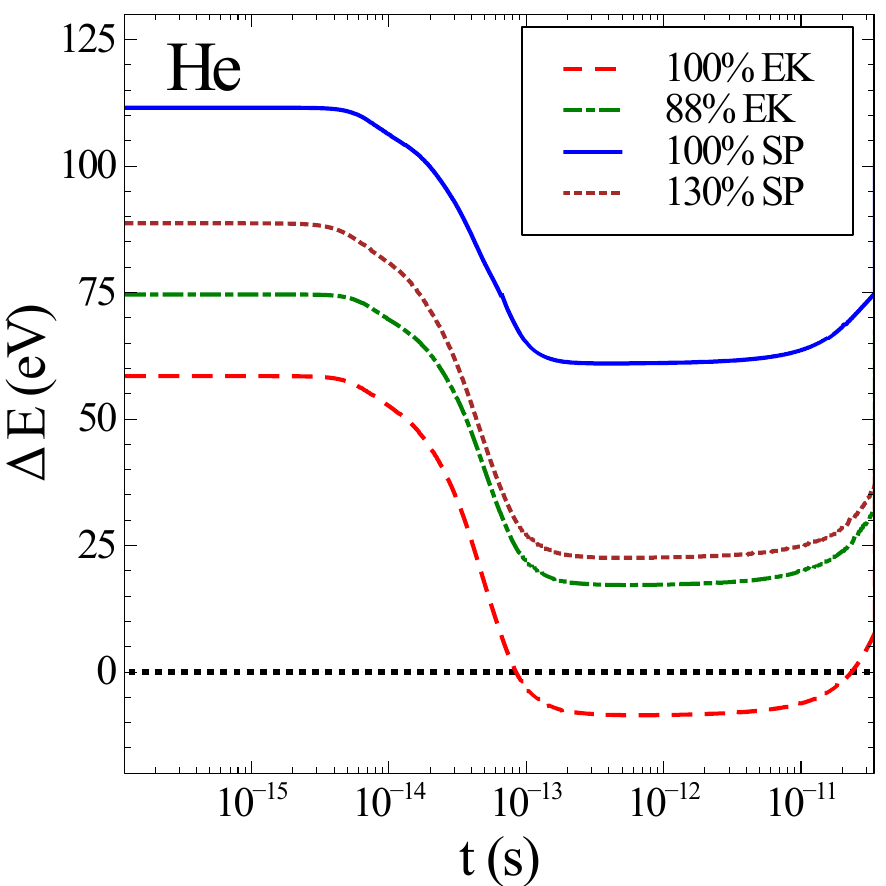}
} 
\subfloat[\label{fig:Continuum_Li}]{%
    \includegraphics[width=0.49\linewidth]{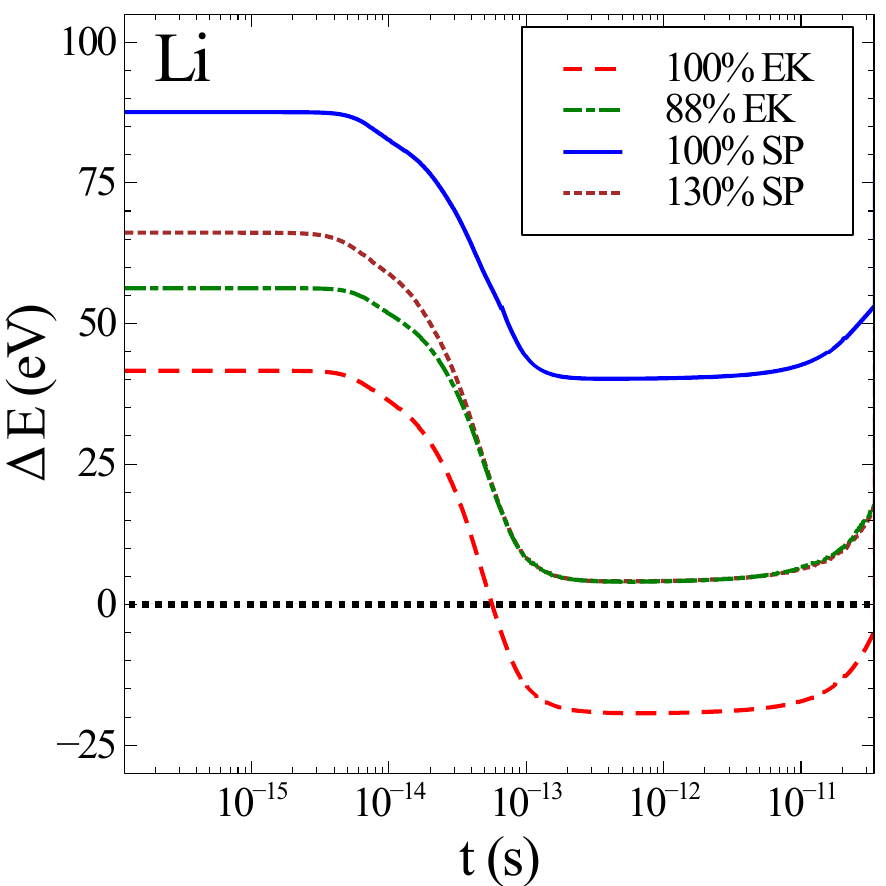}
} 
\caption{\label{fig:IPD_Level}Time-evolution of the level of the continuum edge with respect to the M-shell of He-like (\ref{fig:Continuum_He}) and Li-like (\ref{fig:Continuum_Li}) ions, for the different \IPD models mentioned in the text. In this picture, when the continuum energy is positive, it means that the M-shell is \textit{bound} and when it is negative, the M-shell becomes \textit{free}. Note how, while for the usual \SP and \EK models the M-shell is respectively bound or free for both ion species (when the temperature is sufficiently high), for the case of 88\% \EK and 130\% \SP, the M-shell is barely bound for both charge states and much closer to the continuum edge.}
\end{figure*}

The main difference between these \IPD models can be seen in Figs. \ref{fig:Continuum_He} and \ref{fig:Continuum_Li}, where we show the ionization energy of the M-shell for He- and Li-like ions as a function of time. The \SP model predicts the M-shell to be always bound, with a binding energy around $\sim\unit[100]{eV}$, which is of the same order as the electron energy. This means that a large proportion of the free electrons are able to collisionally ionize these states, losing part of their thermal energy. In contrast, in the \EK model the M-shell becomes completely free when the plasma heats up, so collisional ionization loses are somewhat reduced. Also shown in Fig. \ref{fig:IPD_Level} are the results of scaled \SP and \EK models, such that the M-shell remains bound (in the bound-free picture of atomic kinetics codes), but with a lower ionization energy, which increases the rate of collisional ionization, thus increasing the energy losses and reducing the overall plasma temperature.

\subsection{Spectrum}

Having used SCFLY to determine the temperature evolution of the plasma, the more detailed atomic kinetics were solved using an iterative Saha-Boltzmann \LTE code which treats Ne-like to Be-like ions using configuration averaged levels, while explicitly including the configurations and fine structure levels of Li-like to H-like ions. The energy of the different atomic states and the transition probabilities were obtained from the \LANL atomic codes \cite{LANLCode}. 

The particular region of the spectrum in which we are interested is around the Helium-like resonance line, along with its associated satellites, that is, the region between $1340$ and \unit[1370]{eV} shown in Fig. \ref{fig:FullSpectrum}. 
Since the targets used in this experiment were \unit[54]{nm}-thick, the peak optical depth of the plasma in the spectral range of the \hea emission (determined by the resonance line) is $\tau<0.2$ \cite{preston2017}.  All of the radiation in this region, between 1340  and 1370 eV, is produced by radiative  transitions from the three superconfigurations (110)-(200), (120)-(210), and (111)-(201).

As noted in the introduction, for the solid-density plasmas produced here, the electron temperatures are such that very few K-shell holes are produced thermally for the He and Li ionisation stages,  compared with the number that are produced due to photoionisation by the incident \FEL radiation.  Indeed, during the \FEL pulse, the photoionisation production of such inner core holes exceeds that due to thermal collisional processes by about two orders of magnitude.  However, and importantly for our analysis here, the electron density of the system is so high, and thus electron collisional processes so fast, that the superconfigurations (and also configurations and fine structure levels within them) are extremely close to \LTE, apart from the deviation in the population of the K-shell induced by the photoionisation process.  However, importantly, the photoionization due to the \FEL does not significantly disturb the other aspects of \LTE relationships within the system.  This is best illustrated by example.  Consider two superconfigurations that do not have K-shell holes: (210) and (211). Photoionisation by the \FEL would, from these superconfigurations, produce (110) and (111) respectively, and indeed towards the peak of the pulse the fractional populations of ions with such K-shell holes is of order a few percent.  However, the collisional processes are so fast that the ratio of (110) to (111) remains almost identical to that which would pertain in the circumstances of no \FEL, but the same electron temperature (i.e. that given by the Saha-Boltzmann equation at this temperature). Note also, that given that the (111) superconfiguration can undergo Auger decay, collisional processes are also much more important than this effect. 

\begin{figure*}
\subfloat[\label{fig:110vs111_SP100}]{%
    \includegraphics[width=0.49\linewidth]{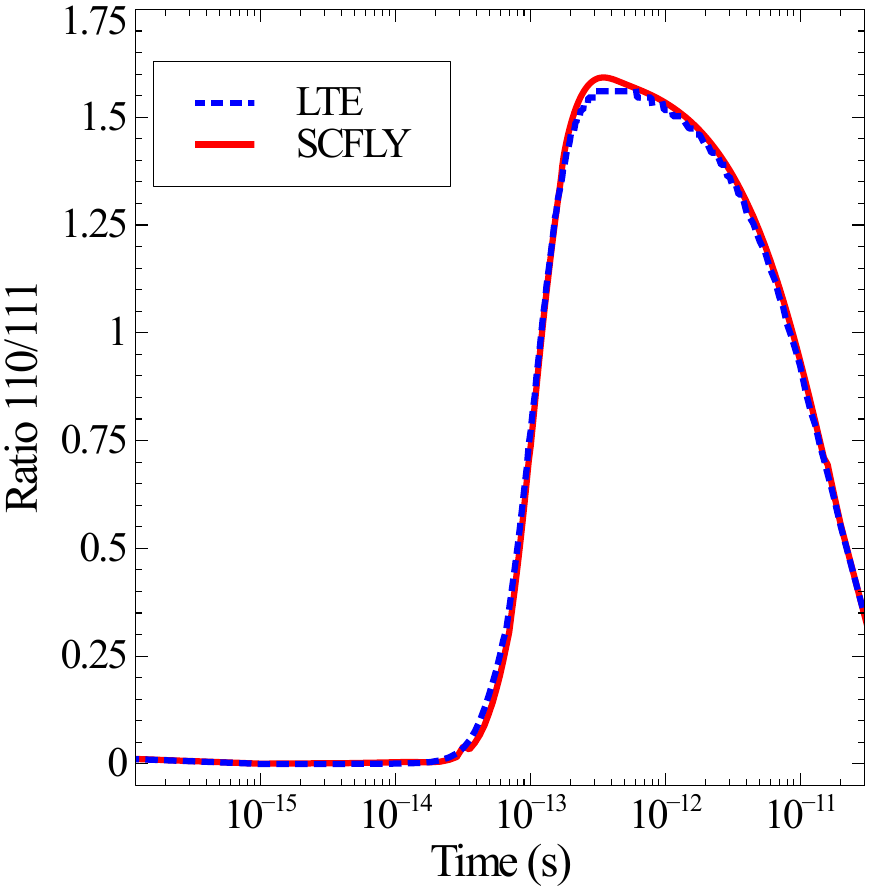}
} 
\subfloat[\label{fig:110and111Pop_SP100}]{%
    \includegraphics[width=0.49\linewidth]{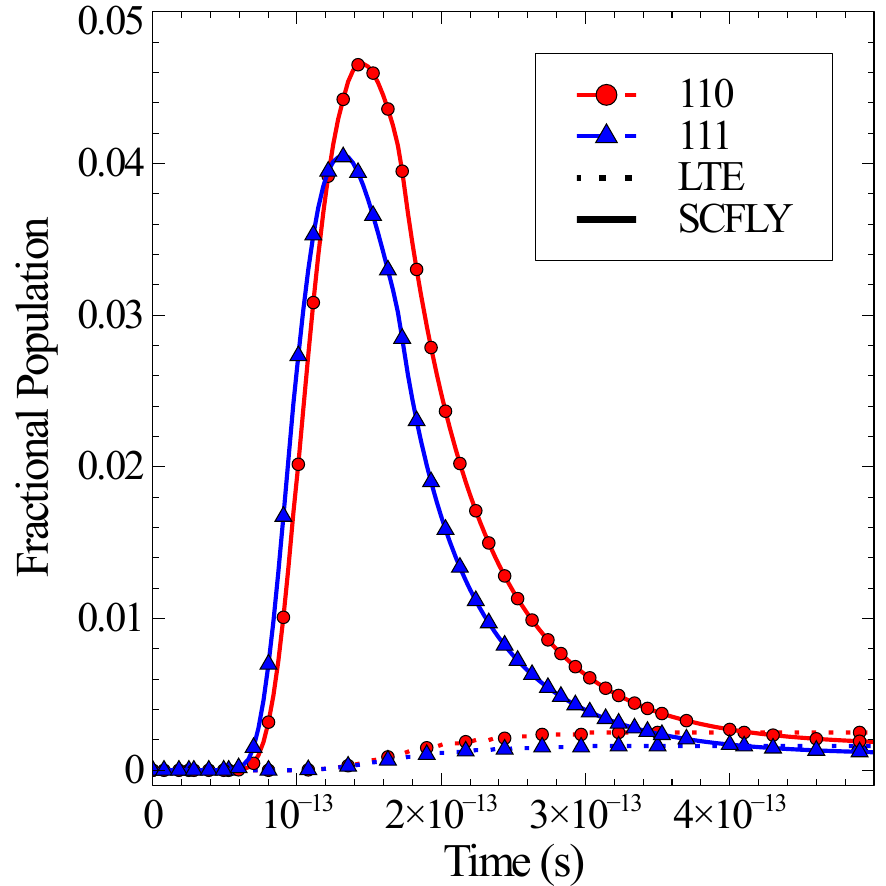}
} 
\caption{\label{fig:110vs111}(\ref{fig:110vs111_SP100}) Population ratio between the 110 and 111 states obtained from SCFLY compared with those obtained by assuming \LTE for the \SP \IPD model as a function of time. (\ref{fig:110and111Pop_SP100}) Fractional population of the He-like 110 states (red circles) and the Li-like 111 satellite states (blue triangles) as a function of time as predicted by SCFLY (solid lines) and pure \LTE (dotted lines).  The symbols are marked every 8 data points to ease the view.}
\end{figure*}

As an example of this we plot the ratio of the populations of the (110) superconfiguration to that of the (111) superconfiguration, as predicted by SCFLY, firstly with the \FEL on and running the simulation in full non-\LTE mode, and then secondly, switching the \FEL off (so no photoionisation occurs), but assuming \LTE, following the time-dependent temperature from the first non-\LTE simulation. As can be seen in Fig. \ref{fig:110vs111_SP100}, the ratio between the populations of the superconfigurations is almost identical at each point in time, and whether the \FEL is on or off.  This is despite the fact that the absolute value of the populations of the superconfigurations is quite different between the non-\LTE calculation including the \FEL, and the one that assumes \LTE: this can be seen in Fig. \ref{fig:110and111Pop_SP100}, where we plot the number density of ions in the (110) and (111) superconfigurations as a function of time for the non-\LTE and \LTE case.  Note that the populations are about two orders of magnitude higher whilst the \FEL is on, due to photoionisation of the K-shell, in the non-\LTE case.

As \LTE between superconfigurations has been shown to be maintained in the presence of an \FEL drive, it is therefore reasonable to use the time-dependent temperature from SCFLY as a basis of the model, assuming \LTE between configurations, and the fine structre levels within them in order to construct a detailed spectrum.  In this limit, as the plasma is optically thin, the intensity emitted at a given time from a given detailed transition depends only on its population and spontaneous emission rate. We use the Saha-Boltzmann solver, with the time-dependent temperature provided by SCFLY, to determine the time-dependent populations of all of the relevant levels, and thus the detailed emission.  For the Helium-like line and its satellites  we included the resonance ($1s2p~\leftidx{^1}P_1\rightarrow 1s^2~\leftidx{^1}S_0$) and intercombination ($1s2p~\leftidx{^3}P_1\rightarrow 1s^2~\leftidx{^1}S_0$) lines from the \hea complex and all Li-like L- and M-shell satellite transitions with energies above \unit[1330]{eV} and with spontaneous emission rates $A$ above $\unit[0.08\times10^{13}]{s^{-1}}$. This corresponds to $13$ L-shell satellite lines and $32$ M-shell satellite lines, which are shown in Table \ref{table:He}. Note that, although the spectral region of interest for this work lies between $1340$ and \unit[1370]{eV}, we set the lower limit of the energy of the lines considered to be \unit[10]{eV} below the region of interest, to include the emission from the wings of the lines. 

\begin{figure*}
\centering
\subfloat[\label{fig:IndLines_SP}]{%
    \includegraphics[width=0.49\linewidth]{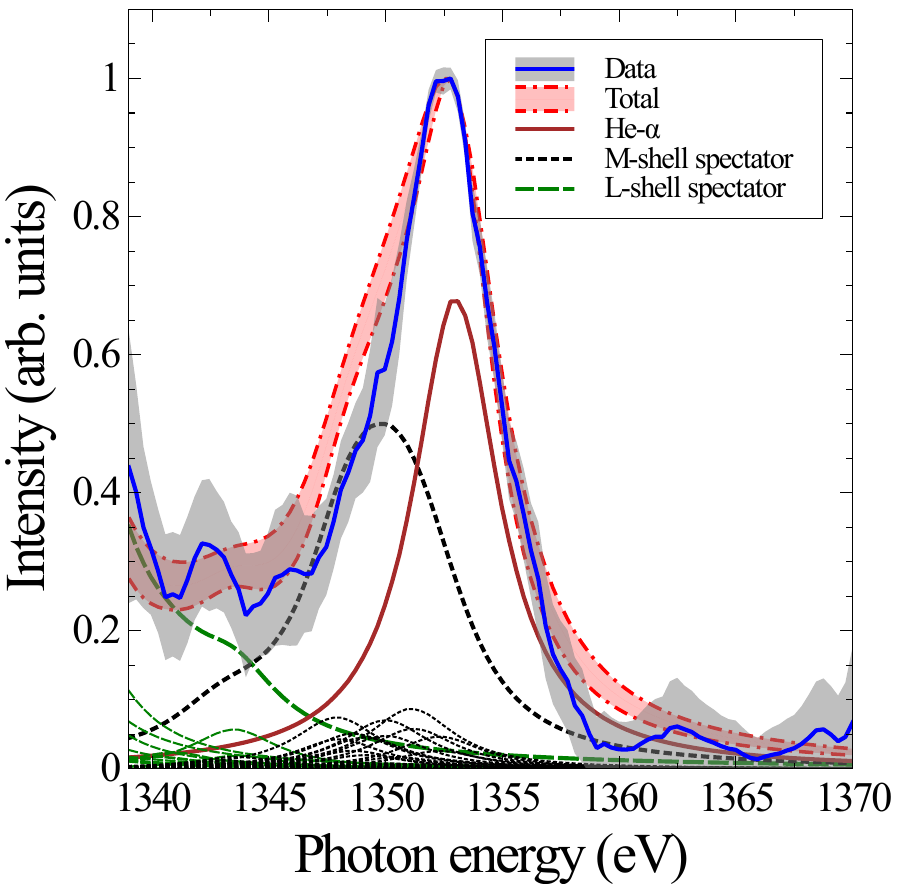}
} 
\subfloat[\label{fig:IndLines_EK}]{%
    \includegraphics[width=0.49\linewidth]{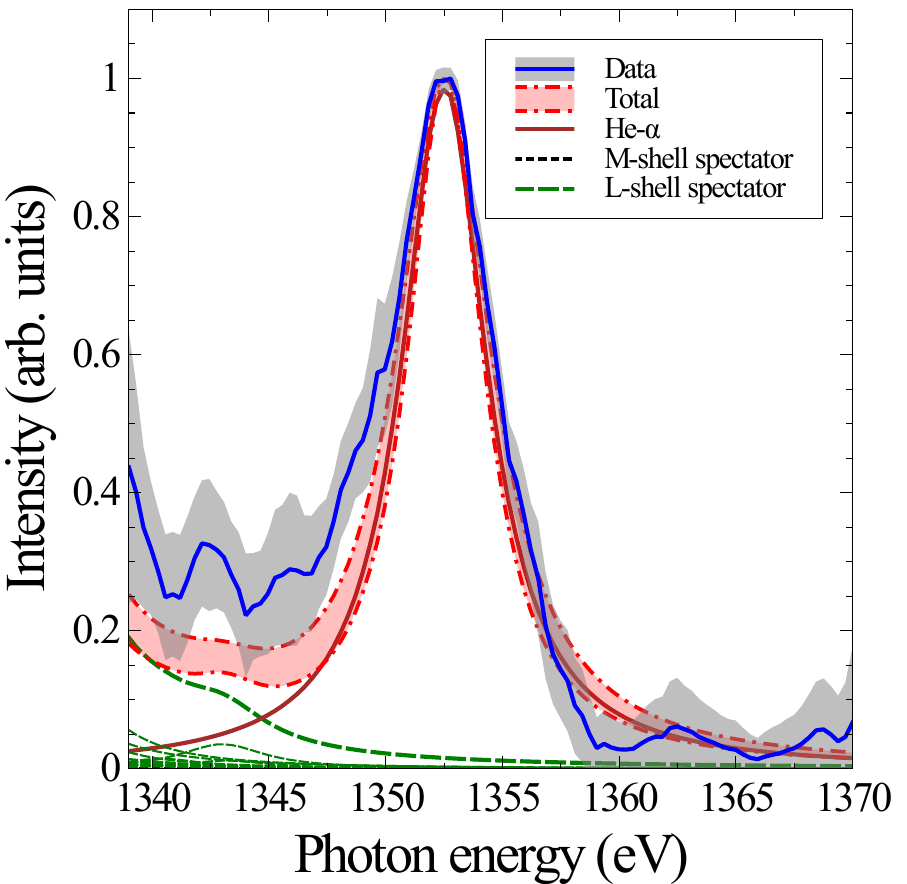}
} 
\caption{\label{fig:IndLines}Shape of the \hea emission obtained by the (\ref{fig:IndLines_SP}) Steward-Pyatt and (\ref{fig:IndLines_EK}) Ecker-Kr\"oll models (red-shaded area, enclosed by dash-dotted lines) compared with the experimental data and the associated $1\sigma$ uncertainty (solid blue line and surrounding grey shaded area respectively). Below the main line, the total contribution from the satellite lines with an M or an L-shell spectator electron, as well as that of the \hea emission are indicated with the dotted black, dashed green and solid brown lines respectively. The thinner dotted black and dashed green lines correspond to the emission from individual fine structure states.}
\end{figure*}

The intensity of each transition is distributed across its lineshape. Given that, for the duration of the emission, the ions are static, the contribution of the Doppler effect to the broadening of the line can be considered negligible. Moreover, during the emission the ions are fixed to their lattice position, and therefore the average ionic microfield felt by a given emitter is zero.\footnote{This effect was verified using the Stark code SIMULA \cite{gigosos2014}. The ions were allowed to shift slightly from their lattice position. No significant mixing was observed for any of the transitions specified here (He$\alpha$, L-shell or M-shell satellites).} Additionally, given the high spectral resolution of the system, the instrumental broadening does not contribute significantly to the lineshape. By studying the mechanisms affecting line broadening in the experimental conditions, using the code ALICE \cite{hill2018}, we observed that, under these conditions, the lineshape is mostly determined by the rate of collisional ionization and recombination, and thus, the lineshape is well modelled by a Lorentzian curve. The experimental width of the isolated \hea resonance line was determined by fitting the high-energy side of the line, where no satellite emission is present. The obtained FWHM of the Lorentzian was $\Delta E = \unit[4.4\pm0.5]{eV}$. We observed that the shape of the total spectrum was not strongly dependent on the width of the individual satellite lines -given the numerous satellite transitions, the shape of the individual features is lost. For this reason, and given that the rate of collisional recombination to the M-shell for a He-like ion is the same as the rate of collisional ionization of an M-shell electron for a Li-like ion, for simplicity we considered all the individual transitions to have the same Lorentzian width as the \hea line. The lineshapes are then modelled as
\begin{equation}
    I(E) = \frac{I_{\text{total}}}{\pi}\frac{\Delta E/2}{(E-E_0)^2+(\Delta E/2)^2},
\end{equation}
where $I_\text{total}$ is the total intensity of a transition defined by the populations and spontaneous rate,  and $E_0$ is the transition energy.

\begin{center}
\begin{table*}[t]
\caption{\label{table:He}List of He-like transitions and Li-like satellites included in the spectra calculation.}
\begin{tabularx}{\linewidth}{>{\centering\arraybackslash}Y >{\centering\arraybackslash}X >{\centering\arraybackslash}Z >{\centering\arraybackslash}Z >{\centering\arraybackslash}Y}
Type & Transition & Energy (eV) & A-rate ($\unit[10^{13}]{s^{-1}}$) & Gabriel's notation \\
\hline & \\[-2ex]
He-like & $1s2p~^3P_1\rightarrow 1s^2~^1S_0$& 1343.1 & $1.21\times10^{-3}$ & y \\
He-like & $1s2p~^1P_1\rightarrow 1s^2~^1S_0$& 1352.5 & 2.0469 & w \\
L-shell satellite & $1s2p^2~^2D_{3/2}\rightarrow 1s^22p^1~^2P_{3/2}$& 1331.3 &  0.0499 & l \\
L-shell satellite & $1s2p^2~^2D_{5/2}\rightarrow 1s^22p^1~^2P_{3/2}$& 1331.3 & 0.9596 & j \\
L-shell satellite & $1s2p^2~^2D_{3/2}\rightarrow 1s^22p^1~^2P_{1/2}$& 1331.8 & 0.9270 & k \\
L-shell satellite & $1s2p^2~^2P_{1/2}\rightarrow 1s^22p^1~^2P_{3/2}$& 1333.4 & 0.9147 & c \\
L-shell satellite & $1s2p^2~^2P_{3/2}\rightarrow 1s^22p^1~^2P_{3/2}$& 1333.8 & 2.6167 & a \\
L-shell satellite & $1s2p^2~^2P_{1/2}\rightarrow 1s^22p^1~^2P_{1/2}$& 1333.8 & 2.0883 & d  \\
L-shell satellite & $1s2p^2~^2P_{3/2}\rightarrow 1s^22p^1~^2P_{1/2}$& 1334.3 & 0.3683 & b  \\
L-shell satellite & $1s2s2p~^2P_{1/2}\rightarrow 1s^22s^1~^2S_{1/2}$& 1335.3 & 1.7592 & r  \\
L-shell satellite & $1s2s2p~^2P_{3/2}\rightarrow 1s^22s^1~^2S_{1/2}$& 1335.6 & 1.8214 & q \\
L-shell satellite & $1s2s2p~^2P_{1/2}\rightarrow 1s^22s^1~^2S_{1/2}$& 1339.9 & 0.2026 & t \\
L-shell satellite & $1s2s2p~^2P_{3/2}\rightarrow 1s^22s^1~^2S_{1/2}$& 1340.4 & 0.1406 & s \\
L-shell satellite & $1s2p^2~^2S_{1/2}\rightarrow 1s^22p^1~^2P_{3/2}$& 1343.0 & 0.7842 & m \\
L-shell satellite & $1s2p^2~^2S_{1/2}\rightarrow 1s^22p^1~^2P_{1/2}$& 1343.2 & 0.2636 & n \\
M-shell satellite & $1s^12p^13p^1~(^3P)^2P_{3/2} \rightarrow 1s^23p^1~^2P_{3/2}$ & 1337.4 & 0.0807 & \\
M-shell satellite & $1s^12p^13p^1~(^3P)^2P_{1/2} \rightarrow 1s^23p^1~^2P_{1/2}$ & 1337.5 & 0.0854 & \\
M-shell satellite & $1s^12s^13d^1~(^1S)^2D_{5/2} \rightarrow 1s^23p^2~^2P_{3/2}$ & 1341.9 & 0.2530 & \\
M-shell satellite & $1s^12p^13p^1~(^3P)^2D_{5/2} \rightarrow 1s^23p^1~^2P_{3/2}$ & 1341.9 & 0.1543 & \\
M-shell satellite & $1s^12s^13d^1~(^1S)^2D_{3/2} \rightarrow 1s^23p^2~^2P_{1/2}$ & 1342.0 & 0.2760 & \\
M-shell satellite & $1s^12p^13d^1~(^3P)^2F_{5/2} \rightarrow 1s^23d^1~^2D_{3/2}$ & 1342.6 & 0.3080 & \\
M-shell satellite & $1s^12p^13d^1~(^3P)^2F_{7/2} \rightarrow 1s^23d^1~^2D_{5/2}$ & 1343.1 & 0.3324 & \\
M-shell satellite & $1s^12p^13p^1~(^3P)^2S_{1/2} \rightarrow 1s^23p^1~^2P_{3/2}$ & 1343.8 & 0.3058 & \\
M-shell satellite & $1s^12p^13p^1~(^3P)^2S_{1/2} \rightarrow 1s^23p^1~^2P_{1/2}$ & 1344.0 & 0.0910 & \\
M-shell satellite & $1s^12p^13s^1~(^3P)^2P_{1/2} \rightarrow 1s^23s^1~^2S_{1/2}$ & 1344.9 & 0.2500 & \\
M-shell satellite & $1s^12p^13s^1~(^3P)^2P_{3/2} \rightarrow 1s^23s^1~^2S_{1/2}$ & 1345.4 & 0.1850 & \\
M-shell satellite & $1s^12p^13p^1~(^1P)^2D_{5/2} \rightarrow 1s^23p^1~^2P_{3/2}$ & 1347.4 & 1.8410 & \\
M-shell satellite & $1s^12p^13p^1~(^1P)^2D_{3/2} \rightarrow 1s^23p^1~^2P_{3/2}$ & 1347.4 & 0.2166 & \\
M-shell satellite & $1s^12p^13p^1~(^1P)^2D_{3/2} \rightarrow 1s^23p^1~^2P_{1/2}$ & 1347.5 & 1.5855 & \\
M-shell satellite & $1s^12p^13s^1~(^1P)^2P_{1/2} \rightarrow 1s^23s^1~^2S_{1/2}$ & 1348.1 & 1.8705 & \\
M-shell satellite & $1s^12p^13s^1~(^1P)^2P_{3/2} \rightarrow 1s^23s^1~^2S_{1/2}$ & 1348.1 & 1.8226 & \\
M-shell satellite & $1s^12p^13p^1~(^1P)^2P_{1/2} \rightarrow 1s^23p^1~^2P_{3/2}$ & 1348.4 & 0.4731 & \\
M-shell satellite & $1s^12p^13p^1~(^1P)^2P_{1/2} \rightarrow 1s^23p^1~^2P_{1/2}$ & 1348.6 & 1.4666 & \\
M-shell satellite & $1s^12p^13p^1~(^1P)^2P_{3/2} \rightarrow 1s^23p^1~^2P_{3/2}$ & 1348.6 & 1.6772 & \\
M-shell satellite & $1s^12p^13d^1~(^1P)^2D_{3/2} \rightarrow 1s^23d^1~^2D_{5/2}$ & 1349.4 & 0.1720 & \\
M-shell satellite & $1s^12p^13d^1~(^1P)^2D_{3/2} \rightarrow 1s^23d^1~^2D_{3/2}$ & 1349.5 & 1.8304 & \\
M-shell satellite & $1s^12p^13d^1~(^1P)^2D_{5/2} \rightarrow 1s^23d^1~^2D_{5/2}$ & 1349.5 & 1.7970 & \\
M-shell satellite & $1s^12p^13d^1~(^1P)^2D_{5/2} \rightarrow 1s^23d^1~^2D_{3/2}$ & 1349.5 & 0.2028 & \\
M-shell satellite & $1s^12p^13p^1~(^1P)^2S_{1/2} \rightarrow 1s^23p^1~^2P_{3/2}$ & 1350.0 & 1.2453 & \\
M-shell satellite & $1s^12p^13p^1~(^1P)^2S_{1/2} \rightarrow 1s^23p^1~^2P_{1/2}$ & 1350.1 & 0.4075 & \\
M-shell satellite & $1s^12p^13d^1~(^1P)^2F_{7/2} \rightarrow 1s^23d^1~^2D_{5/2}$ & 1350.6 & 1.6623 & \\
M-shell satellite & $1s^12p^13d^1~(^1P)^2F_{5/2} \rightarrow 1s^23d^1~^2D_{5/2}$ & 1350.7 & 0.1690 & \\
M-shell satellite & $1s^12p^13d^1~(^1P)^2F_{5/2} \rightarrow 1s^23d^1~^2D_{3/2}$ & 1350.7 & 1.4779 & \\
M-shell satellite & $1s^12p^13d^1~(^1P)^2P_{1/2} \rightarrow 1s^23d^1~^2D_{3/2}$ & 1351.8 & 1.9741 & \\
M-shell satellite & $1s^12p^13d^1~(^1P)^2P_{3/2} \rightarrow 1s^23d^1~^2D_{5/2}$ & 1351.8 & 1.7840 & \\
M-shell satellite & $1s^12p^13d^1~(^3P)^2P_{3/2} \rightarrow 1s^23s^1~^2S_{1/2}$ & 1351.9 & 0.1761 & \\
M-shell satellite & $1s^12p^13d^1~(^3P)^2P_{1/2} \rightarrow 1s^23s^1~^2S_{1/2}$ & 1352.4 & 0.1594 & \\
\end{tabularx}
\end{table*}
\end{center}

\subsection{Effect of the focal spot}
\label{SS:focal}

In order to take into account the spatial distribution of the focussed \FEL x-rays, for each pulse we model  25 intensity bins spanning six orders of magnitude, as detailed by Ciricosta \textit{et al.} \cite{ciricosta2016}. The intensity of the $n$-th bin was obtained as
\begin{equation}
    I_n(\omega,t) = I_0(\omega,t)\cdot e^{-2(n\cdot0.1)^2}.
    \label{eqn:LaserFluence}
\end{equation}
It is worth mentioning that not all the bins contribute to the \hea emission, since for $n\gtrsim 11$ the temperature is not sufficiently high to ionize the the plasma up to neither Li nor He-like state. This spatial distribution was measured experimentally by the imprint method, as described in more detail in~\cite{ciricosta2016}. 
The full spectrum was then obtained as a sum of the time-integrated spectra calculated for each intensity bin, weighted by the fluence scan of the laser spot.
 
\begin{figure*}
\centering
\subfloat[\label{fig:Full_line}]{%
    \includegraphics[width=0.49\linewidth]{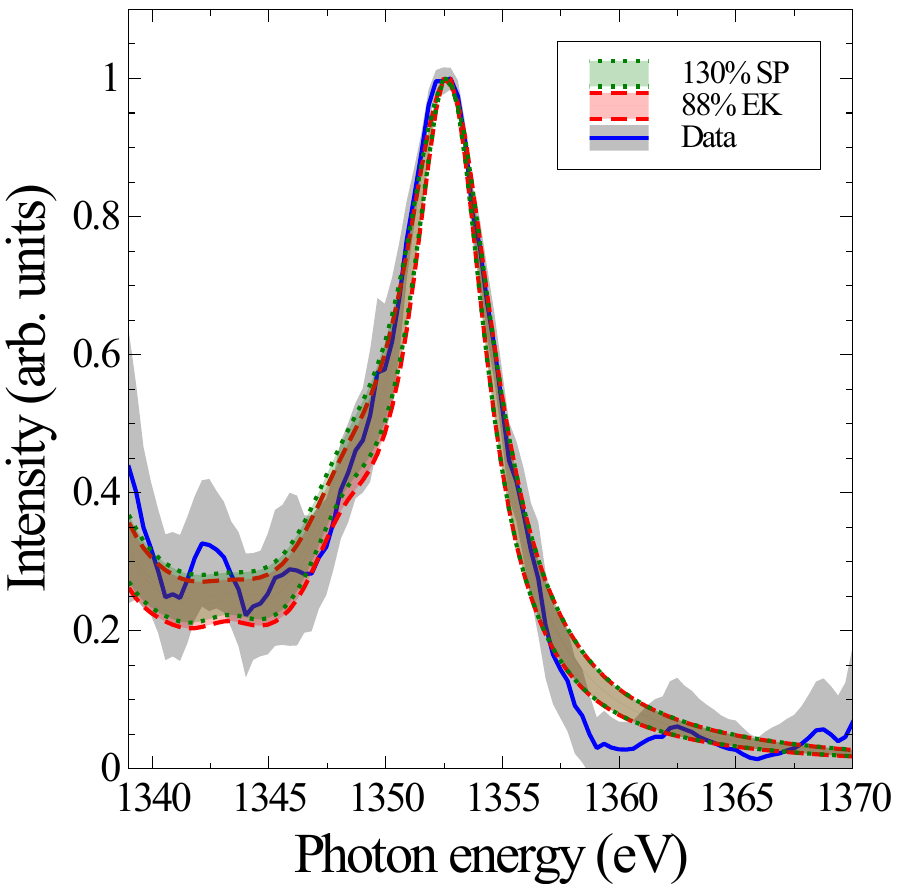}
} 
\subfloat[\label{fig:IndLines_SP130}]{%
    \includegraphics[width=0.49\linewidth]{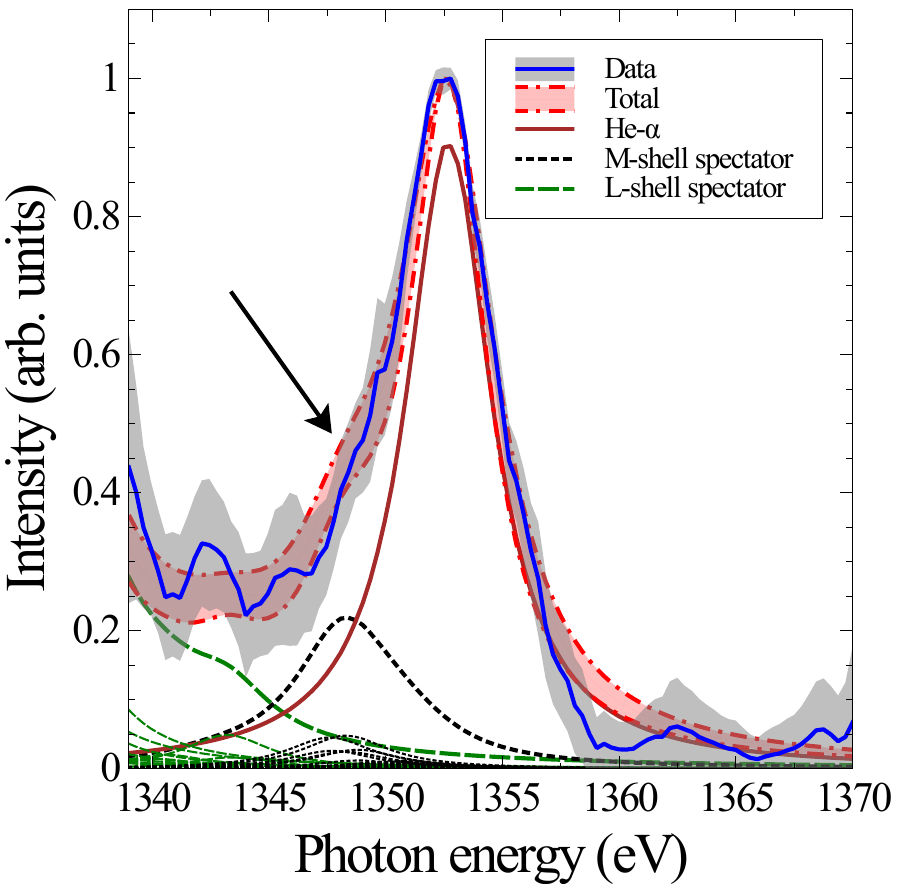}
} 
\caption{\label{fig:Correct_IPDs}(\ref{fig:Full_line}) Comparison of the shapes of the \hea emission obtained for both 130\% of the \SP model and 88\% of the \EK model (green-dotted and red-dashed shaded regions respectively) compared with the experimental data. The width of the error regions corresponds to the uncertainty in the line widths. (\ref{fig:IndLines_SP130}) Line profile obtained using the \SP \IPD model scaled to 130\% (red-shaded area enclosed by dash-dotted lines) compared to the experimental data (solid blue lines and grey area). The contribution from individual lines is shown following the same colour convention as Fig. \ref{fig:IndLines}. The black arrow marks the shoulder-like feature caused by the M-shell satellite emission.}
\end{figure*}

\section{Results}

Figure \ref{fig:IndLines} shows both the experimental data, and simulated spectra, for the Helium-like emission and associated Li-like satellites. We show the simulated spectra of the \hea region for both the \SP (Fig. \ref{fig:IndLines_SP}) and the \EK (Fig. \ref{fig:IndLines_EK}) \IPD models. The solid brown line corresponds to the contribution from the \hea emission, while the dashed green and dotted-black lines correspond to the collective emission from the Li-like satellites with an  L- and M-shell spectator electron respectively (the thinner dashed green and dotted black lines correspond to each individual satellite transition). The red-shaded band enclosed by dash-dotted lines corresponds to the total spectrum, with the width of the band corresponding to the uncertainty introduced by the error in the fit to the width of the Lorentzian. When comparing with the experimental data (shown as a solid blue line surrounded by a grey-shaded area corresponding to a 1$\sigma$ uncertainty), it  can be seen that the \SP model overestimates the intensity of the M-shell satellites, thus predicting a \hea feature $\sim\unit[2]{eV}$ wider than the experimental result. The opposite happens with the \EK model, where there is almost no M-shell satellite contribution, and therefore the line appears narrower than observed. 

It is worth noting that the solid-brown line labeled He-$\alpha$ in the figures includes the contribution from the resonance line $1s2p~\leftidx{^1}P_1\rightarrow 1s^2~\leftidx{^1}S_0$ (labeled $w$ following Gabriel's notation \cite{gabriel1972}), centered at $E_w=\unit[1352.5]{eV}$; and the intercombination line $1s2p~\leftidx{^3}P_1\rightarrow 1s^2~\leftidx{^1}S_0$ ($y$), centered at $E_y = \unit[1343.1]{eV}$. The contribution of the intercombination line, however, can barely be resolved, since its intensity is $\sim 0.2\%$ that of the resonance line.  Whilst the intercombination line is a well-known intense feature of spectra from plasmas produced by irradiation with optical lasers, it is almost completely absent here.  This is due to the very high electron densities present in these experiments, which contrast with the critical electron densities in optical experiments.  With optically produced laser plasmas, which are far from \LTE, significant population can build up in the $1s2p~^3P_1$  state, giving rise to a large intercombination line intensity, despite the very low transition rate to the lower $1s^2~^1S_0$ level.  However, here the collisional effects cause a much lower \LTE-dictated population of $1s2p~^3P_1$, leading to almost complete absence of the transition.

From these figures, it seems clear that the intensity of the M-shell satellites seen experimentally cannot be reproduced by either the full \SP or \EK models, at least for the \FEL intensity used.  However, if we consider the intensity of these satellites to indicate in some way where the \IPD should actually lie, it would be somewhere between the two.
We thus reran the SCFLY and Saha-Boltzmann solver for the two \IPD models, but
changed the value of $C$ within them from unity, while keeping the functional form of the model the same (see Eqns. \ref{eqn:EK} and \ref{eqn:SP}). 

We performed a $\chi^2$ minimization fit, where the width of the lines was allowed to vary within the measured uncertainty ($\unit[4.4\pm0.5]{eV}$). The best fits were obtained for $C_{EK}=0.88\pm0.03$ and $C_{SP}=1.30\pm0.04$, cases for which both models predict the M-shell to remain just \textit{bound} for the whole duration of the emission (see Figs. \ref{fig:Continuum_He} and \ref{fig:Continuum_Li} respectively).

\begin{figure*}
\subfloat[\label{fig:Diff_Int_EK}]{%
   \includegraphics[width=0.49\linewidth]{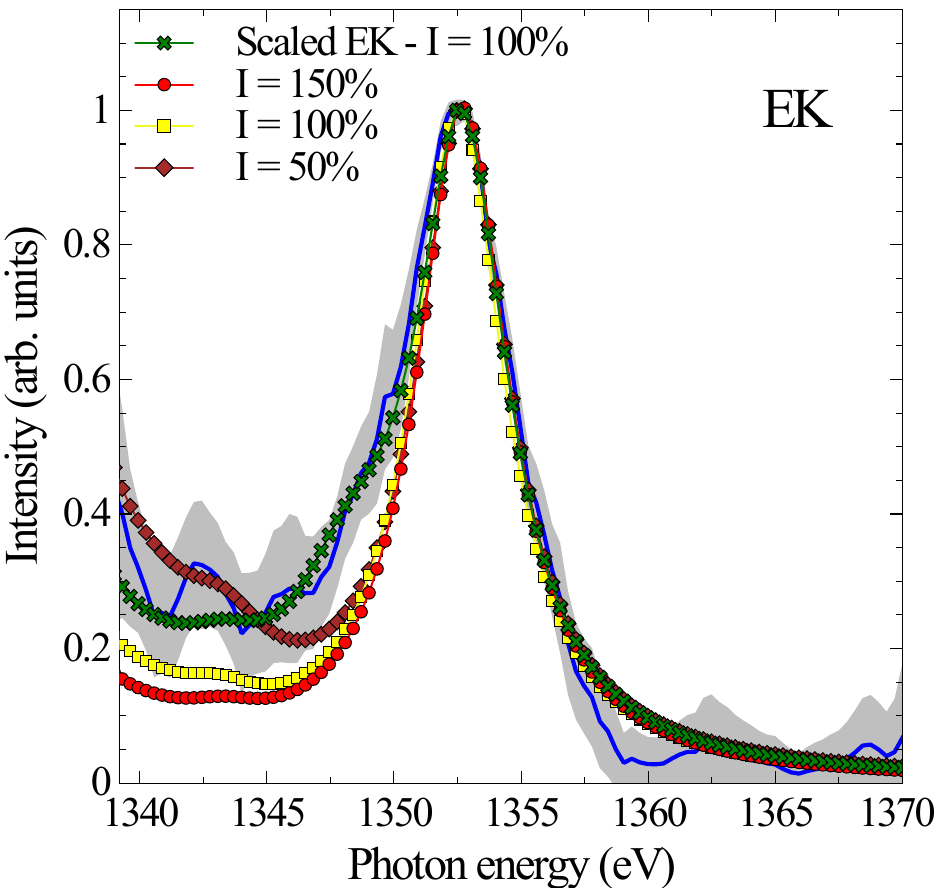}
} 
\subfloat[\label{fig:Diff_Int_SP}]{%
    \includegraphics[width=0.49\linewidth]{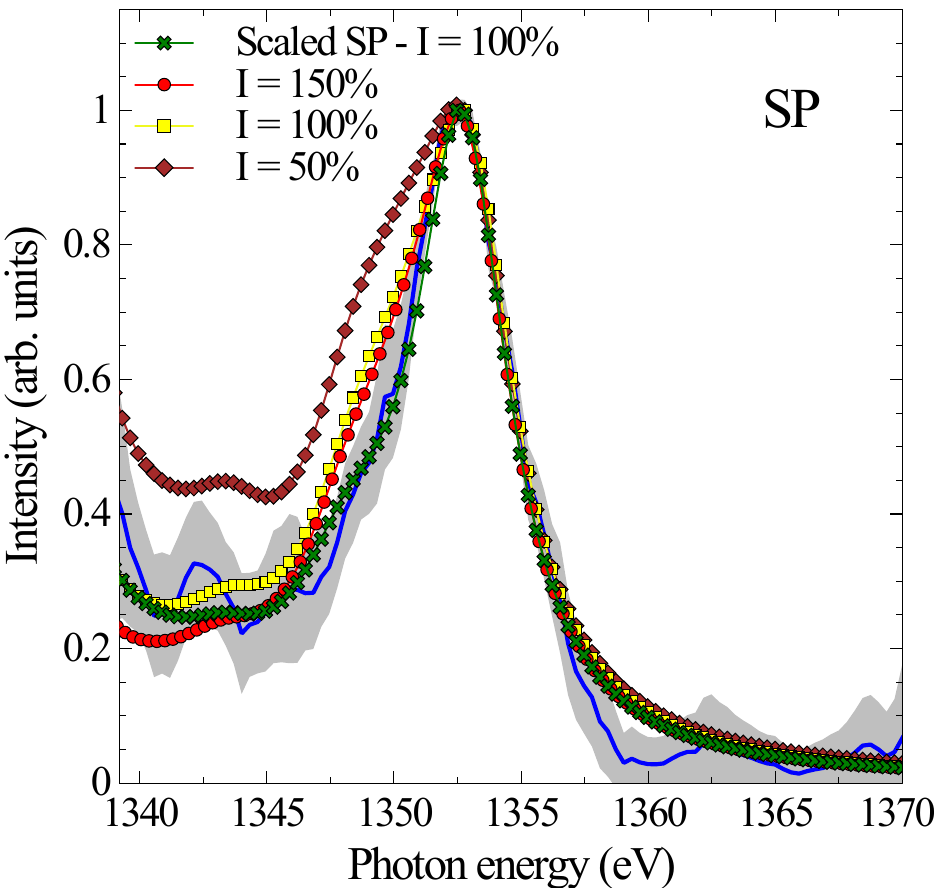}
} 
\caption{\label{fig:Diff_Int}Spectra obtained for the \EK (\ref{fig:Diff_Int_EK}) and \SP (\ref{fig:Diff_Int_SP}) \IPD models for the nominal laser intensity (yellow squares), and the results obtained by modifying the laser intensity to 150\% (red circles) and 50\% the nominal value (brown diamonds). The results obtained by scaling the \IPD models, as presented in Fig. \ref{fig:Correct_IPDs}, are also shown for comparison (green crosses).}
\end{figure*}

The resulting spectra are shown in Fig. \ref{fig:Correct_IPDs}, where \ref{fig:Full_line} shows the total spectra for both cases and Fig. \ref{fig:IndLines_SP130} corresponds to the individual contribution of each line for the 130\% \SP \IPD model, following the same color convention as for Fig. \ref{fig:IndLines}. It can be seen that the emission from the M-shell satellites is reponsible for the shape of the low-energy wing of the line (where it creates a shoulder-like feature that we are indicating with an arrow), whereas the \hea emission is the only contribution on the high-energy end.

In the context of the recent \DFT calculations by Gawne \textit{et al.} \cite{gawne2023}, it is encouraging that to obtain the best fit to the satellite spectra we need to invoke an \IPD value that lies between the \SP and \EK limits, as that was precisely the conclusion of that work. Furthermore, the required scaling of both the \EK and \SP models to match the experimental data is in excellent agreement with the required scaling to match the \DFT results (Gawne \textit{et al.} found an \IPD value corresponding to 133\% that predicted by the \SP model and 89\% the prediction of the \EK model). However, before stating definitively that this is the case, care should be taken to note that the simulated intensity of the satellites compared with the resonance line will depend on accurate modelling of the temperature of the system, as the satellite intensity is determined by the ratio of the ionisation energy of their upper levels to the temperature. This in turn, as well as any inherent limitations of the model, entails accurately knowing the experimental x-ray \FEL intensity incident upon the target: a figure that is generally quoted to be known within about 30\% \cite{ciricosta2016}. It is therefore important to also investigate the sensitivity of the results to the \FEL intensity on target. 


To this end we ran several simulations with \FEL intensities that differed from those measured experimentally by up to a factor of two lower and up to 50\% higher than the experimental value. As described previously in subsection \ref{SS:focal}, these simulations were integrated over the shape of the focal spot, in order to  be directly comparable with the current results.  The results are shown in Fig. \ref{fig:Diff_Int_EK} for the full \EK model and in Fig. \ref{fig:Diff_Int_SP} for the full \SP model. In these figures, the result for an scaled \IPD level of 88\% of the \EK model and 130\% of the \SP model respectively is included for comparison. As expected, in the spectral region of the M-shell satellites, the predicted line width differs very little in the case of the \EK model, as the M-shell electrons are not bound in any event. The main changes to the spectrum appear in the region of the L-shell satellites, which become more intense with respect to the \hea emission as the laser intensity is decreased.  

However, in the case of using the full \SP model, a reduction in the laser intensity does lead to an increase in satellite intensity predicting an even wider line. On the other hand, an increase in the intensity slightly reduces the satellite contribution from both M- and L-shell satellites. We find that increasing the laser intensity by $50\%$ results in a line width that is still $\sim\unit[1]{eV}$ wider than the experimental result, while the emission around \unit[1340-1345]{eV} starts deviating from the data as well. Furthermore, looking in more detail we see the shoulder asymmetry mentioned before in the lower energy side to the main \hea peak ($\sim\unit[1348]{eV}$) appearing for the scaled-\IPD models. This feature is not present in the full \SP simulations, but does appear in the data.  We thus conclude that given we believe we know the incident \FEL flux to within about 30\%, the simulations still indicate that the satellite intensities are more consistent with an \IPD model that lies between the \EK and \SP limits once the He and Li-like ion stages are reached.
 
\section{Conclusions}

In this work we have investigated the sensitivity of the intensity of M and L-shell satellites to the \hea emission from a solid-density plasma to the \IPD model used.  By adjusting the \IPD level in an atomic-kinetics code, in conjunction with an \LTE Saha-Boltzmann solver, we find that both the \SP and \EK \IPD models fail to reproduce the experimental spectra, but obtain best agreement with the experimental data by employing a degree of \IPD that lies between these two extremes,  ($88\pm3$\% of \EK and $130\pm4$\% of \SP). These values are in good agreement with recent results obtained using first principles simulations, and  indicate that the M-shell of Li-like and He-like Mg under these conditions lies very close to the continuum edge.  Whilst the intensity of the M-shell satellites does depend on the intensity of the \FEL, our knowledge of the incident intensity, and the observation of asymmetry in the \hea peak, does lead credence to the conclusion that the best fit \IPD lies between the \EK and \SP limits. 

As outlined in the introduction, it should be borne in mind that  for such dense quantum systems the distinction between bound and free states is somewhat artificial, yet in the mode in which current atomic-kinetics calculations are performed, we are often forced to make this division.  The fact that the experimental spectra can be reasonably reproduced within this simple framework, using \IPD values guided by first-principle simulations, gives confidence that they are still of use in the modelling of the spectra of hot dense plasmas.

\section*{Acknowledgements}

T.G., J.S.W. and S.M.V. acknowledge support from AWE via the Oxford Centre for High Energy Density Science (OxCHEDS). S.M.V. acknowledges support from the Royal Society. J.S.W. and S.M.V. acknowledge support from the UK EPSRC under grants EP/P015794/1 and EP/W010097/1. G.P.-C. acknowledges support from Spanish Ministry of Science and Innovation under Research Grants No. PID2019-108764RB-I00 and PID2022-137632OB-I00 from the Spanish Ministry of Science and Innovation. S.M.V. is a Royal Society University Research Fellow. T.B., J.Ch., V.H., L.J., and V.V. appreciate financial support from the Czech Ministry of Education (grant no. LM2023068) and Czech Science Foundation (grant No. 20-08452S).

Use of the Linac Coherent Light Source (LCLS), SLAC National Accelerator Laboratory, is supported by the U.S. Department of Energy, Office of Science, Office of Basic Energy Sciences under Contract No. DE-AC02-76SF00515.

\bibliography{bibliography3}

\end{document}